\documentclass[twocolumn,floatfix, showpacs]{revtex4}

\usepackage[final]{epsfig}
\usepackage{amsmath}
\usepackage{parskip}
 
\begin{document}
 
\title{A comparison of various measures for the average transport coefficient $\hat{q}$}
 
\author{Thorsten Renk}
\email{thorsten.i.renk@jyu.fi}
\affiliation{Department of Physics, P.O. Box 35 FI-40014 University of Jyv\"askyl\"a, Finland}
\affiliation{Helsinki Institute of Physics, P.O. Box 64 FI-00014, University of Helsinki, Finland}
 

\begin{abstract}
Jet quenching, i.e. the suppression of high transverse momentum ($P_T$) hadron production in ultrarelativistic heavy-ion collisions is among the most striking experimental signatures of bulk medium formation. Efforts with the aim of extracting quantitative information about the bulk medium from the measured suppression mainly focus on the extraction of an averaged transport coefficient $\langle \hat{q} \rangle$ as a measure of the medium jet quenching power, with the underlying assumption that  $\langle \hat{q} \rangle$ is a meaningful quantity to make comparisons both among different models and between models and data. In this note, the main uncertainties associated with the extraction of $\hat{q}$ from model fits to data are briefly reviewed before the notion of an average transport coefficient itself is investigated. It is shown in a case study that the choice of a meaningful average is far from unique and that different well-motivated definitions of averages lead not only to differences in the outcome of about a factor four, but also in a different change of $\langle \hat{q} \rangle$ with collision centrality. This casts some doubt on the idea that condensing the information about jet-medium interactions into a single average parameter is a meaningful procedure.
\end{abstract}
 
\maketitle

\section{Introduction}

The suppression of high transverse momentum ($P_T$) hadron production in heavy-ion (A-A) collisions with respect to the scaled expectation value from p-p collisions is experimentally well established from both single hadron observables \cite{PHENIX_R_AA,PHENIX-RP} and hard dihadron correlations \cite{Dijets1,Dijets2}. Theoretical models using well-constrained fluid dynamical simulations of the bulk matter evolution as the background in which high $p_T$ processes take place can describe these observables qualitatively well \cite{Comparison,Dijets}, thus the natural next step is to extract quantitative parameters of the medium which characterize its ability to modify high $P_T$ partons from the experiments.

Since detailed differential observables like the average medium induced parton energy loss $\Delta E$ probability distribution $P(\Delta E)$ are difficult to obtain from single hadron observables as the nuclear suppression factor $R_{AA}$ is insensitive to the functional form of $P(\Delta E)$ \cite{gamma-h,inversion}, the focus in recent times has been on the determination of the averaged transport coefficient $\langle \hat{q} \rangle$ (the average virtuality gain per unit pathlength of a parton travelling in the medium) or an equivalent parameter within various models of energy loss from detailed fits to single hadron and dihadron observables (see e.g. \cite{ZOWW,Armesto} for such attempts).

Unfortunately, there is a history of large differences in the value of $\langle \hat{q} \rangle$ extracted from different models, and some of these differences persist even in a comprehensive analysis of three differet models under the same set of assumptions about the bulk medium evolution \cite{Comparison}. A couple of reasons why different calculations may result in different medium parameters have been identified since:

\begin{itemize}

\item The assumptions about the medium geometry and evolution may not be the same. While for example early computations in the Armesto-Salgado-Wiedemann (ASW) formalism \cite{QuenchingWeights} were done in a static cylindrical geometry and found an averaged $\hat{q} = 5..15$ GeV$^2$/fm, later computations utilized the same formalism in a fluid-dynamical background \cite{Armesto,Jet-Hydro} and found averaged transport coefficients almost an order of magnitude smaller. As results presented in \cite{Jet-Hydro} indicate, even the variation between dynamical medium evolution models which are constrained by the medium induces an uncertinty if $\sim 50$ \% on the extraction of a $\hat{q}$.

\item The model connecting thermodynamical parameters with a value of $\hat{q}$ may not be the same. Common prescriptions in the absence of flow use $\hat{q} \sim s$, $\hat{q} \sim T^3$ or $\hat{q} \sim \epsilon^{3/4}$ where $s,T$ and $\epsilon$ are the entropy density, temperature and energy density as taken from a hydrodynamical description of the medium. While all these relations lead to the same result if the medium can be described as an ideal gas, they result in an additional $\sim 50$ \% uncertainty in the extraction of $\hat{q}$ if e.g. a bag model equation of state is used for the medium \cite{Comparison}.

\item For formalisms derived using a collinear approximation, prescriptions to treat finite kinematics must be used in the application of the formalism to describe data. In doing so, a large systematic uncertainty is introduced by the detailed choice of the finite kinematics prescription \cite{Uncertainties}.

\end{itemize}

\section{Averaging procedures}

However, even if all these issues can be overcome, any extraction of a $\hat{q}$ from an evolving medium eventually faces the question how a meaningful average can be defined. From a theoretical perspective, $\hat{q}$ is a local property of the medium. In the static case, it is tied to local medium thermodynamics, leading to the dependence $\hat{q}(\eta_s,r,\phi,\tau)$ on the complete set of parameters spacetime rapidity $\eta_s$, radius $r$, angle $\phi$ and proper evolution time $\tau$. In a flowing medium, $\hat{q}$ acquires a dependence on the relative orientation of the direction of parton propagation and the flow velocity as \cite{Flow}

\begin{equation}
\label{E-Flow}
\hat{q} = \hat{q}_{static} (\cosh \rho - \sinh \rho \cos\alpha)
\end{equation}

where $\rho$ is the flow rapidity and $\alpha$ the relative angle between flow and parton trajectory. In principle, the value of $\hat{q}$ should also depend on the resolution scale $E$ at which the parton probes the medium \cite{Scale}. Trying to condense the full information contained in $\hat{q}(\eta_s,r,\phi,\tau,\rho,\alpha,E)$ into a single number is a bit like asking a Meteorologist to summarize the weather forecast for a place for the next 48 hours in a single number --- it may not nesessarily be a meaningful exercise. In the formalism comparison study \cite{Comparison}, the problem has been circumvented by comparing every energy loss model for the same hydrodynamical medium and making the single highest value of $\hat{q}$ the standard of comparison. This works as almost the whole parameter dependence of $\hat{q}$ is specified by the hydrodynamical medium, but results in values of $\hat{q}$ which are only meaningful for this particular evolution model and do now allow a meaningful comparison to results obtained in other models for the bulk medium.

In the following, several different averaging procedures are discussed. Note that all computations are done for the same ASW energy loss  formalism and the same 3-d hydrodynamical bulk matter evolution model as described in detail in \cite{Comparison}. In all cases, the model reproduces $R_{AA}(P_T)$ for centeral and non-central collisions. Any differences in the extracted values of $\langle \hat{q}\rangle$ quoted below are thus exclusively caused by the different averaging procedures. 

\subsection{Spacetime averaging}

One possibility to define an average transport coefficient is to make averages over the spacetime evolution of the hydrodynamical medium, given a model for the relation between $\hat{q}$ and thermodynamical parameters. Dropping the flow correction Eq.~(\ref{E-Flow}), a relation like

\begin{equation}
\hat{q}(\eta_s,r,\phi,\tau) = K \cdot \epsilon(\eta_s,r,\phi,\tau)^{3/4}
\end{equation}

can be used to compute the nuclear suppression factor $R_{AA}$ as outlined e.g. in \cite{Comparison,Dijets,Jet-Hydro} with $K$ adjusted such that the data is described. For the average transport coefficient, the expression

\begin{equation}
\label{E-Av}
\langle \hat{q} \rangle = \frac{\int d\eta_s dr d \phi d \tau \hat{q}(\eta_s,r,\phi,\tau)}{\int d\eta_s dr d \phi d \tau}
\end{equation}

where the integration boundaries are given by the thermalization and freeze-out hypersurfaces of the medium. If this is done for central collisions at 200 AGeV using the hydro model of \cite{Hydro}, a value of $\langle \hat{q} \rangle = 0.82$ GeV$^2$/fm is obtained. Note that this result is based on exactly the same ASW formalism described in \cite{Comparison} with a $\hat{q}_{max} = 18.5$ GeV$^2$/fm.

However, this prescription has an unwanted dependence on the choice of the freeze-out conditions. Increasing the freeze-out temperature $T_F$ from 130 MeV (the standard choice in \cite{Hydro}) to 140 MeV does not alter the value of $K$ obtained in a fit to $R_{AA}$, but the value of $\langle \hat{q} \rangle$ changes to $1.02$ GeV$^2$/fm. 
The physical reason for this is of course that hard partons tend to escape from the medium long before bulk freeze-out occurs, thus whatever happens at late times in the evolution does not influence hard parton production, yet the late time behaviour of the hydrodynamics still enters the averaging procedure Eq.~(\ref{E-Av}). In particular, this implies that two models in which the interaction between hard partons and the medium happens in exactly the same way may still lead to different values of $\langle \hat{q} \rangle$ if their late-time dynamics is different.

If one cuts the averaging after a finite time with the argument that partons only see the evolution before this cut, the extracted average transport coefficient is significantly larger. Assuming all partons escape by the time of 5 fm/c increases $\langle \hat{q} \rangle = 1.8$ GeV$^2$/fm, changing the cut to 3 fm/c yields 2.33 GeV$^2$/fm. While such a cut removes the artificial smallness of the extracted transport coefficient due to the late time evolution of the medium, $\langle \hat{q} \rangle$ is then dependent on the relatively arbitrary choice of a timescale. Thus, spacetime averages do not provide a satisfactory solution to all aspects of the problem.

\subsection{Path averaging}

Since the weakness of the spacetime averaging approach is that the fact that hard partons leave the medium after a finite time is not taken into account, one may think of averaging along a single parton path as

\begin{equation}
\label{E-PathAv}
\langle \hat{q} \rangle_{path} = \frac{\int d \zeta \hat{q} (\zeta)}{\int d \zeta}
\end{equation}

where $\zeta$ is a variable describing the spacetime trajectory of the parton through the medium and the integration boundaries are again given by the intersection of the parton trajectory with the thermalization and freeze-out hypersurfaces. If one assumes that the initial vertices of the hard processes responsible for parton production scale as the number of binary collision with the overlap

\begin{equation}
\label{E-Profile}
P(x_0,y_0) = \frac{T_{A}({\bf r_0 + b/2}) T_A(\bf r_0 - b/2)}{T_{AA}({\bf b})},
\end{equation}

where the thickness function is given in terms of Woods-Saxon the nuclear density
$\rho_{A}({\bf r},z)$ as $T_{A}({\bf r})=\int dz \rho_{A}({\bf r},z)$, the average transport coefficient can be computed by averaging Eq.~(\ref{E-PathAv}) over all possible initial positions $(x_0,y_0)$ and angles with a weight given by Eq.~(\ref{E-Profile}). If this is done in the model outlined above, one finds  $\langle \hat{q} \rangle = 2.2$ GeV$^2$/fm. This is consistent with the spacetime average with a cut in $\tau$ between 3 and 4 fm/c and indicates that the bulk of hard partons leaves the medium within this timescale. As expected, the value obtained in the path-based averaging is significantly higher than the spacetime average over the whole medium, as the path averaging shifts weight to the dense medium center and to early times.

However, once one tracks individual parton trajectories, one may pose the question if one should take into account the fact that propagating gluons carry a different color charge and couple by a factor 9/4 more strongly to the medium. If the relative weight of quarks and gluons is obtained in a leanding order perturbative Quantum Chromodynamics Calculation, a value of  $\langle \hat{q} \rangle = 3.4$ GeV$^2$/fm is obtained with a small $P_T$ dependence caused by the change in relative quark and gluon yield at different $P_T$.

But one may also argue that averaging over all parton trajectories is not suitable to make a comparison with the data, because the measurement of the nuclear suppression factor reflects only partons which, after energy loss and fragmentation, have produced a sufficiently high $P_T$ hadron. If one supplements the path-based averaging with a condition that the leading hadron after fragmentation must be found above a certain $P_T$ scale, the average weights change somewhat and the result acquires a scale dependence (which is in practice small for RHIC kinematics but may be noticeable at the LHC). For a scale of 6 GeV, $\langle \hat{q} \rangle = 2.1$ GeV$^2$/fm is found. This number is smaller than the one found when averaging using all paths, as the trigger condition removes weight from the dense regions of the medium where partons are absorbed. In extreme cases, this may be a problem: If one assumes a threshold core-halo scenario in which partons are either absorbed by a dense core or escape undisturbed from a halo, the path-based averaging with trigger condition would find $\langle \hat{q} \rangle = 0$ since only partons from the halo without medium effect are observed.

\subsection{Centrality dependence}

\begin{figure}
\epsfig{file=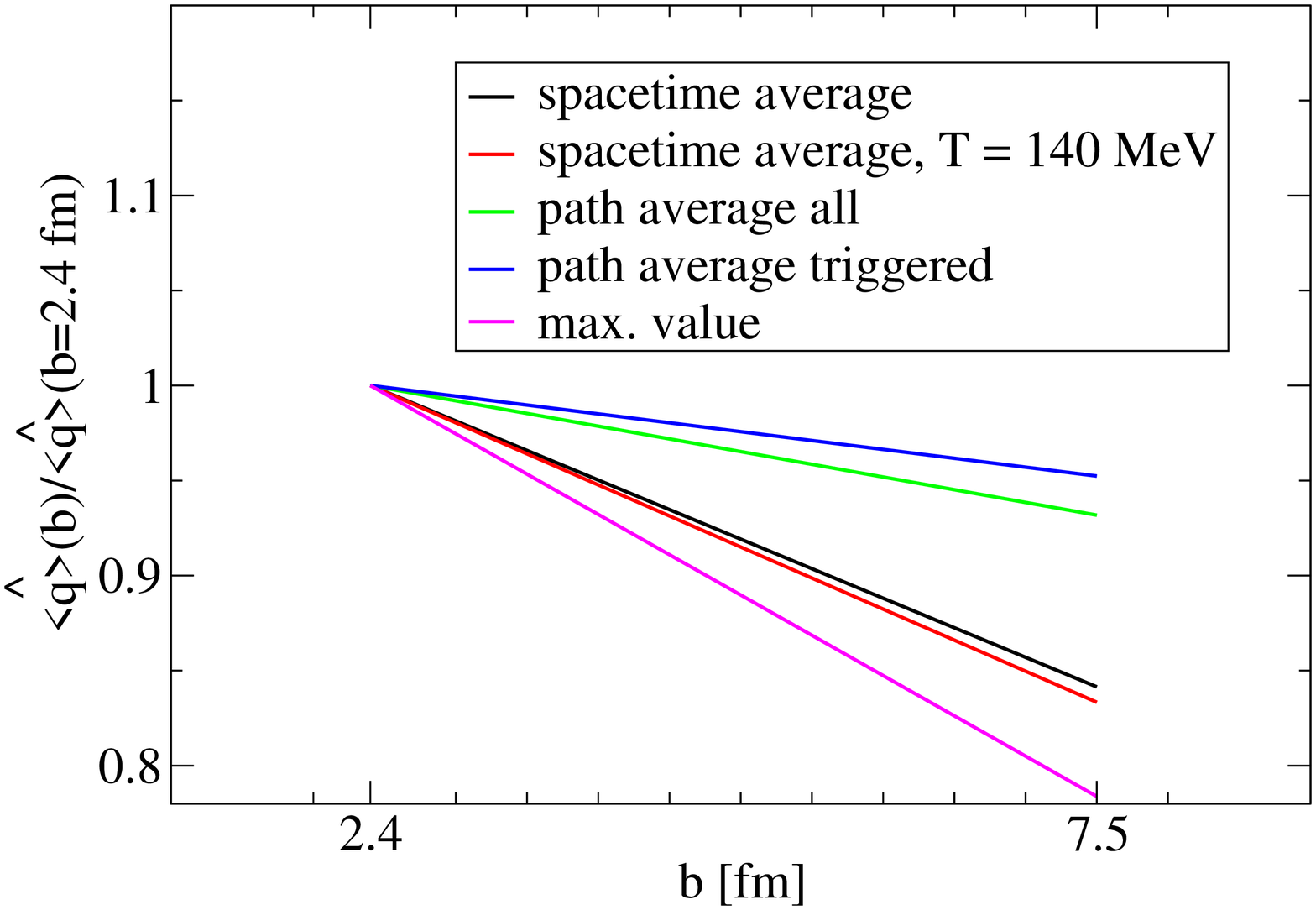, width=7.8cm}
\caption{\label{F-Centrality}(Color online) }
\end{figure}

Conceptually, one wants to define $\langle \hat{q} \rangle$ as a quantity which is extracted from and related to a measured quantity such as $R_{AA}$, but on the other hand characterizes the produced medium. Therefore, any $\langle \hat{q} \rangle$ should reflect a change of the medium properties in a characteristic way. This can for example be tested by studying the relative change of $\langle \hat{q} \rangle$ when going to non-central collisions in the hydrodynamical evolution. Fig.~\ref{F-Centrality} shows the average transport coefficient, normalized by its value for $b=2.4$ fm, for the various averaging procedures described above along with the maximal $\hat{q}$ measure as used in \cite{Comparison} when the medium is changed to non-central collisions. 

It is obvious from the figure that the relative change of $\langle \hat{q} \rangle$ with centrality does not agree among different averaging procedures. The underlying reason is again that different regions within the medium volume receive different weight factors in different averaging procedures.

\section{Discussion}

One can roughly classify the various measures of $\langle \hat{q} \rangle$ according to how much they focus on medium properties as opposed to the measurement process in which the medium is probed. Spacetime averages of the hydro medium incorporate the full information about the medium, but none about the physics underlying $R_{AA}$. Consequently, they provide a good standard of comparison among models of the bulk medium, but given the constraint that models should describe the experimental data for $R_{AA}$, different bulk models may lead to large variations in the extraction of $\langle \hat{q} \rangle$ as the averaging probe spacetime volumes of the medium which are never probed by hard processes.

In contrast, averages based on actual parton propagation and energy loss tie different bulk evolution models much closer to the measurement. But for these averaging procedures, there are dense regions to which the averaging process is blind as almost no partons escape. This implies that two models required to describe the measured $R_{AA}$ may  result in a similar  $\langle \hat{q} \rangle$ in spite of the fact that the models, e.g. in terms of central energy densities, are very different. Thus, constraining a path-averaged  $\langle \hat{q} \rangle$ does not necessarily constrain the underlying bulk evolution model well.

Given these findings, it would appear that a more differential approach to characterize medium properties is needed to do quantitative measurements of medium properties.

\begin{acknowledgments}
 
Discussions with Roy Lacey are greatfully acknowkedged as the starting point of this work. This work was supported by an Academy Research Fellowship from the Finnish Academy and from Academy Project 115262. 
 
\end{acknowledgments}

\end{document}